\documentclass[aps,prl,reprint,superscriptaddress]{revtex4-2}

\usepackage{amsfonts}
\usepackage{amssymb}
\usepackage{amsmath}
\usepackage{graphicx}
\usepackage{epsfig}
\usepackage{array}
\usepackage{braket}
\usepackage{multirow}
\usepackage[table,xcdraw]{xcolor}
\usepackage[colorlinks=true, citecolor=blue, urlcolor=blue, linkcolor=red]{hyperref}

\begin{document}

\title{Non-Hermitian theory of valley excitons in two-dimensional
semiconductors}
\author{Qiutong Wang}
\altaffiliation{These authors contributed equally to this work.} 
\author{Ci Li}
\altaffiliation{These authors contributed equally to this work.}
\author{Qingjun Tong}
\email{tongqj@hnu.edu.cn}
\affiliation{School of Physics and Electronics, Hunan University, Changsha
410082, China}

\begin{abstract}
Electron-hole exchange interaction in two-dimensional transition metal
dichalcogenides is extremely strong due to the dimension reduction, which promises valley-superposed excitonic states
with linearly polarized optical emissions. However, strong circular
polarization reflecting valley-polarized excitonic states is commonly
observed in helicity-resolved optical experiments. Here we present a
non-Hermitian theory of valley excitons by incorporating optical pumping and
intrinsic decay, which unveils an anomalous valley-polarized excitonic state
with elliptically polarized optical emission. This novel state arises
from the non-Hermiticity induced parity-time ($\mathcal{PT}$)-symmetry breaking, which impedes the experimental observation of intervalley excitonic coherence effect. At large excitonic center-of-mass momenta, the $\mathcal{PT}$-symmetry is restored and the excitonic states recover their valley
coherence. Interestingly, the linear polarization directions in optical
emissions from these valley-superposed excitonic states are
non-orthogonal and even become parallel at exceptional points. Our
non-Hermitian theory also predicts a non-zero Berry curvature for valley
excitons, which admits a topological excitonic Hall transport beyond the
Hermitian predictions.
\end{abstract}

\maketitle

\begin{figure}[tbp]
\begin{center}
\includegraphics[width=0.475\textwidth]{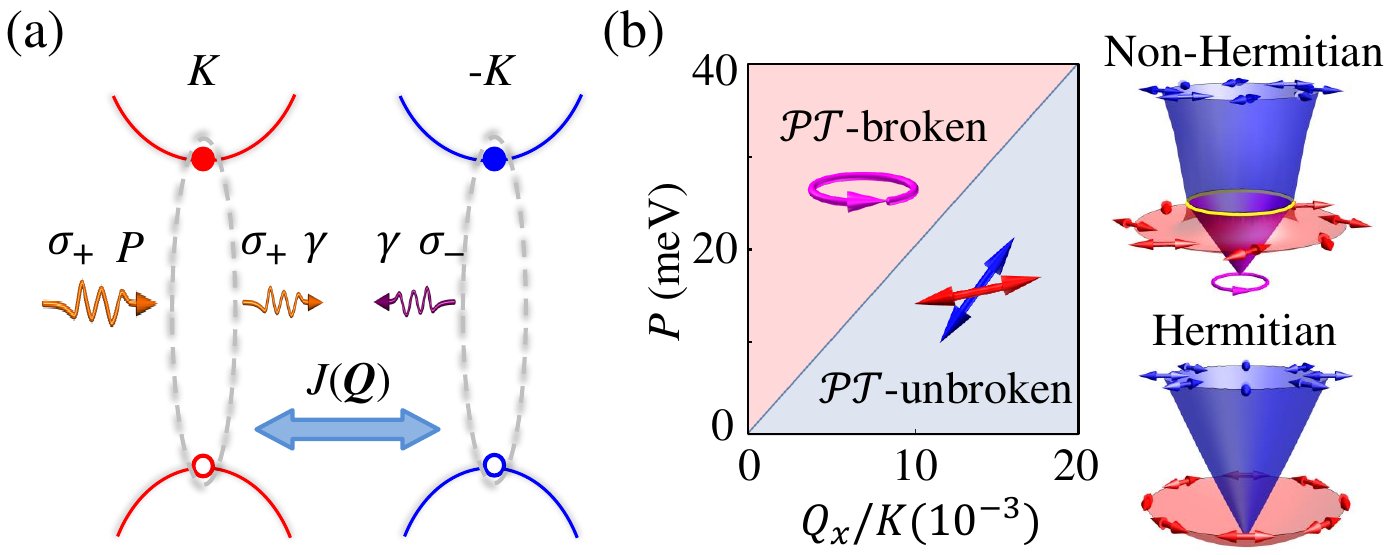}
\end{center}
\caption{(color online) (a) Schematic of a non-Hermitian valley excitonic
model. A $\protect\sigma _{+}$ light pumping $K$ valley with strength $P$,
which is coupled to $-K$ valley via intervalley exchange coupling $J(%
\boldsymbol{Q})$. Both valleys suffer decay $\protect\gamma $. (b) Left:
Phase diagram of non-Hermitian valley excitons as functions of pumping $P$
and excitonic center-of-mass momentum $Q_{x}$. Optical emission is elliptically
polarized in the $\mathcal{PT}$-symmetry broken regime, while linearly
polarized in the unbroken one. Right: Typical excitonic band dispersion in
Hermitian and its real part in non-Hermitian case with the yellow circle marking the exceptional ring. The emission
properties are indicated.}
\label{fig1}
\end{figure}

\textcolor{blue}{Introduction}-Recent years have witnessed rapid progress in
exploring excitonic physics in monolayer transition metal dichalcogenides
(TMDs) and their van der Waals assembles \cite%
{Hei,Yao1,Shan1,Zhang,Urb,Xu1,Li,FW,Shan}. These materials feature a visible
range bandgap located at Brillouin zone corners ($\pm K$
points), introducing a valley degree of freedom for information
encoding \cite{Xu,Yao}. Due to enhanced Coulomb interaction in
two-dimensional (2D) geometry, tightly bound
Wannier excitons dominate optical response in these materials, which can
be endowed with the valley pseudospin through polarization selection rules 
\cite{Xu,Yao,Niu3,Yu}. Another characteristic feature of monolayer TMDs is
the existence of a strong electron-hole exchange interaction, which
couples the valley pseudospin of an exciton to its center-of-mass
motion \cite{Yu,Yu1,Mac,Lou}. This interaction combined with momentum
scattering is responsible for the experimentally observed depolarization
dynamics of valley excitons \cite{Wu,Urb1,Hao}. Furthermore, this
pseudospin-orbital interaction couples coherently the two valleys and
results in valley-superposed excitonic states, which predicts linearly
polarized optical emissions \cite{Yu,Yu1,Mac,Lou}, as illustrated in
Fig. \ref{fig1}(b). However, in helicity-resolved photoluminescence
measurements, strong circular polarization reflecting valley-polarized
excitonic states is commonly observed \cite{Xu,Cui,Hei1,Cao,Sallen,Jones}. Although extrinsic perturbation-induced spectrum broadening in practical experiments \cite{Moody,Ajayi,Cadiz} may weaken the intervalley excitonic coherence effect, a deep understanding from intrinsic mechanism is still demanding.

As quasi-particles, excitons require optical pumping for their
formation and, meanwhile, suffer electron-hole recombination \cite{Haug}, which
renders the excitonic system an intrinsically non-Hermitian nature.
Non-Hermitian physics has recently emerged as one of the most active fields
that intensively studied in diverse artificial quantum systems \cite%
{Chr,Ued1,Kun,ZBL,Ma,Lee}. Because of the non-Hermiticity, eigenvalues of
the Hamiltonian are generically complex, whose imaginary parts are
associated with either quasi-particle decay or growth. An important progress
in exploring non-Hermitian physics is the introduction of a parity-time ($%
\mathcal{PT}$)-symmetry \cite{Ben,Ben1} and its associated breaking
transition at the exceptional points where eigenstates coalesce 
\cite{Hes}, leading to a plethora of novel phenomena without Hermitian
counterparts \cite%
{Zhang1,Hodaei,Wiersig,ZhongLiu,Chen,Ued,Sat,Wang,Du,Son1,Miri,Yang,Li2023}. Bringing non-Hermiticity into valley excitons in 2D semiconductors would not only give an alternative dynamical understanding on their intriguing optical phenomena, but also
open up a low-dimensional solid-state avenue to explore non-Hermitian
physics.

In this letter, we present a non-Hermitian theory of valley excitons in 2D
TMDs via incorporating a circularly polarized optical pumping and an
intrinsic decay. We find that the non-Hermiticity
introduces an exceptional ring located at excitonic center-of-mass momentum space. In
particular, the $\mathcal{PT}$-symmetry is broken at small excitonic
momenta, which results in a novel valley-polarized excitonic state with
elliptically polarized optical emission. This anomalous valley polarization serves as an intrinsic mechanism that obstructs the experimental observation of intervalley excitonic coherence effect. The $\mathcal{%
PT}$-symmetry is restored at large momenta, where the optical
emission becomes linearly polarized. Different from the Hermitian scenarios,
where the linear polarization directions from the two excitonic states are
orthogonal, the ones in non-Hermitian case are rotated
relatively towards each other and even become parallel at the exceptional ring.
Finally, we show that the non-Hermiticity also gives rise to a nontrivial
Berry curvature for valley excitons, which enables an anomalous excitonic Hall effect beyond the Hermitian
predictions.

\begin{figure}[tbp]
\begin{center}
\includegraphics[width=0.475\textwidth]{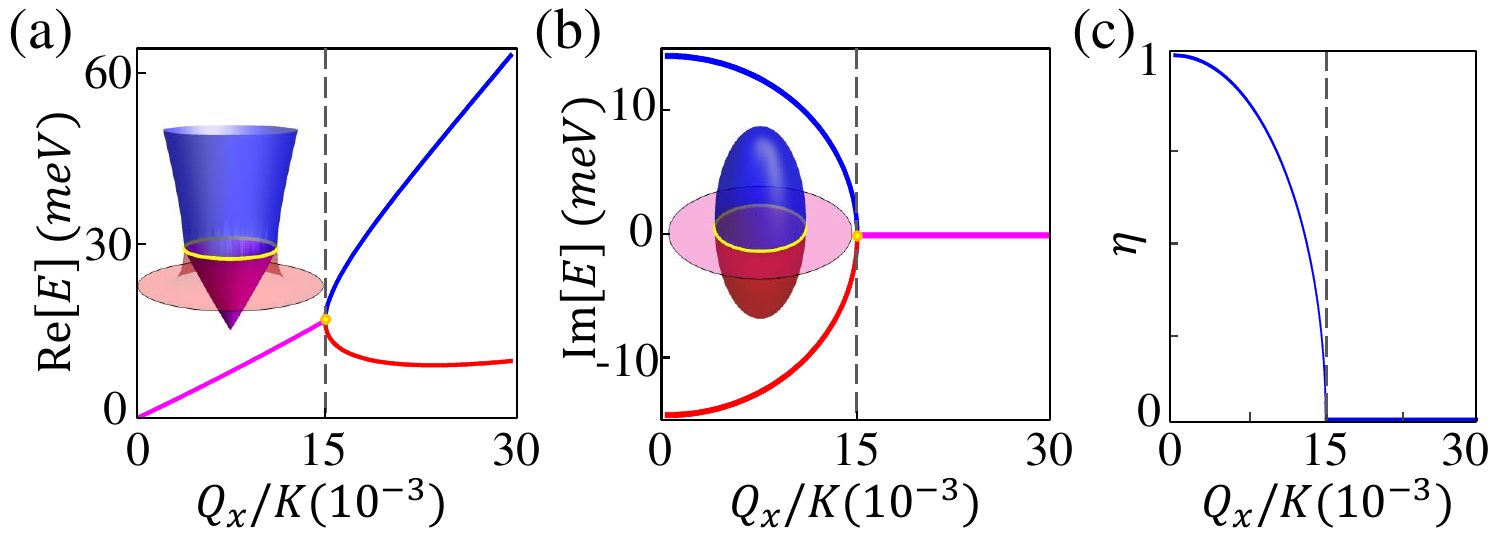}
\end{center}
\caption{(color online) (a) Real and (b) imaginary parts of $E_{\pm }
$ in Eq. (\protect\ref{ep}) as a function of excitonic momentum $Q_{x}$. The
insets show their distribution in 2D momentum space, with the
yellow circles marking the exceptional ring. (c) Valley polarization $\eta$
(quantifying circular polarization in optical experiments) as a function
of $Q_{x}$ for eigenstate $\left\vert u_{+}^{R}\right\rangle $ in Eq. (%
\protect\ref{v}). The black vertical line marks the exceptional points in $%
Q_{x}$ direction. $P=30\,\mathrm{meV}$ is used.}
\label{fig2}
\end{figure}

\textcolor{blue}{Non-Hermitian valley excitons and valley polarization}-With
a circularly polarized optical pumping and intrinsic decay (c.f. Fig. \ref{fig1}(a)), the valley
excitons in 2D TMDs are described by a non-Hermitian Hamiltonian 
\begin{eqnarray}
H_{\mathrm{ex}}\left( \boldsymbol{Q}\right)  &=&\frac{\hbar ^{2}Q^{2}}{2M_{0}%
}+\frac{iP}{2}-i\gamma +J\frac{Q}{K}  \label{H} \\
&&+\left( 
\begin{array}{cc}
iP/2 & J\frac{Q}{K}e^{-2i\varphi } \\ 
J\frac{Q}{K}e^{2i\varphi } & -iP/2%
\end{array}%
\right) ,  \notag
\end{eqnarray}%
acting on the excitonic valley basis $\left\{ \left\vert K\right\rangle
,\left\vert -K\right\rangle \right\} $, $J\sim 1\,\mathrm{eV}$ is the
exchange coupling constant extracted from \textit{first-principles} calculations \cite{Yu1}, $M_{0}$ is the excitonic mass which
approximately equals to the free electron mass \cite{Urb}, $\boldsymbol{Q}%
=Q\left( \cos \varphi ,\sin \varphi \right) $ is its center-of-mass momentum, $K=4\pi
/3a$ with $a$ being the lattice constant of monolayer TMDs ($a\approx 3.31\,%
\mathrm{\mathring{A}}$ of WSe$_{2}$ \cite{Zhu2011} is used here), $\gamma $ is the exciton decay rate (typically on the order of meV \cite{Xu}), and $P
$ is its formation rate controlled by excitation power of a helicity-dependent light ($\sigma _{+}$ used here). The Hermitian counterpart of this model well describes lowest-energy bright A exciton and also B exciton with spin-orbit interaction included, consistent with the simulation from \textit{ab initio} GW-Bethe-Salpeter equation method \cite{Lou}. The effect of other excitonic states can be regarded as an additional decay or pumping reservoir \cite{Supp}. In the following, we consider the steady-state case with $P=2\gamma$, when the Hamiltonian is $\mathcal{PT%
}$-symmetric (see Supplemental Material \cite{Supp}). We note that the non-steady-state case only
introduces a global decay or growth that does not affect the non-Hermitian
physics we revealed here, because the Hamiltonian still has a passive $\mathcal{PT%
	}$-symmetry \cite{Guo2009}. The eigenvalues of this $\mathcal{PT%
}$-symmetric non-Hermitian Hamiltonian
are%
\begin{equation}
E _{\pm }=\frac{\hbar ^{2}Q^{2}}{2M_{0}}+J\frac{Q}{K}\pm \frac{%
\sqrt{4J^{2}Q^{2}-P^{2}K^{2}}}{2K},  \label{ep}
\end{equation}%
which has an exceptional ring at $Q_{0}=\frac{PK}{2J}$, separating the whole $Q$-$P$
parameter space into a $\mathcal{PT}$-symmetry broken phase (at small $Q$
and strong $P)$ and unbroken one (at large $Q$ and
weak $P$) (c.f. Fig. \ref{fig1}(b)). This exceptional ring does not
rely on the steady-state condition, but is determined by the pumping rate $P$%
. Figs. \ref{fig2}(a) and \ref{fig2}(b) plot the real and imaginary parts of
the two eigenvalues as a function of excitonic momentum. As a general feature
of $\mathcal{PT}$-symmetric non-Hermitian system, the energy spectra are
real in the $\mathcal{PT}$-symmetry unbroken regime, while complex in
the broken one. 

We first make a comparison between the excitonic band dispersions for the
Hermitian and non-Hermitian valley excitons, focusing on the real part
in the latter case (see Figs. \ref{fig1}(b) and \ref{fig2}(a)). In the
Hermitian case ($P=0$), the energy spectra reduce to $E _{\pm
}^{H}=\frac{\hbar ^{2}Q^{2}}{2M_{0}}+J\frac{Q}{K}\pm J\frac{Q}{K}$. At small 
$Q$, the excitonic bands have a lower parabolic dispersion and an upper
nonanalytic $v$-shape dispersion with a Fermi velocity of $\frac{2J}{\hbar K}$ \cite{Yu,Yu1,Mac,Lou}. In contrast, in the $\mathcal{PT}$-symmetry broken regime of the non-Hermitian
case, the real parts of the two energy bands are degenerate and have a
nonanalytic $v$-shape dispersion similar as the upper band in the
Hermitian case, but with an approximately half of its Fermi velocity, i.e., $%
\frac{J}{\hbar K}$. This non-Hermiticity induced band degeneracy, together with extrinsic perturbation-induced spectrum broadening \cite{Moody,Ajayi,Cadiz}, may obstruct the observation of excitonic band splitting in practical experiments.  In the $\mathcal{PT}$-symmetric regime, the two bands
split as $\propto \sqrt{Q-Q_{0}}$ near the exceptional ring, which is a typical
spectrum behavior near the exceptional transition points \cite{Hes} and is qualitatively different
from the $\propto Q$ splitting behavior in the Hermitian case. This enhanced splitting, as being widely explored for sensor applications \cite{Wiersig,ZhongLiu,Chen}, results
in an anomalous drop in \textrm{Re}$\left[ E _{-}\right] $ when crossing the exceptional ring from $\mathcal{PT}$-symmetry broken
regime to unbroken one. 

\begin{figure*}[tbp]
\begin{center}
\includegraphics[width=0.75\textwidth]{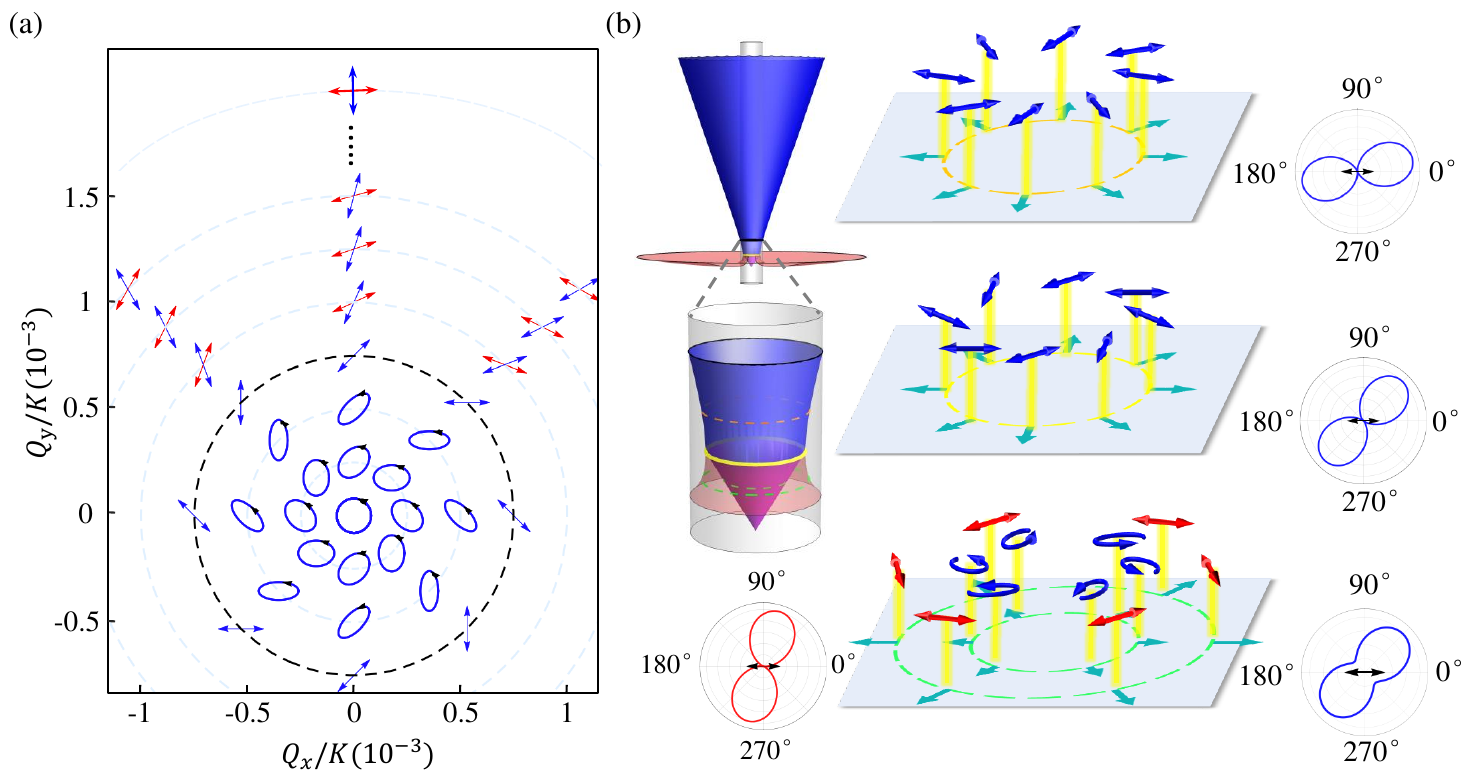}
\end{center}
\caption{(color online) (a) Momentum resolved optical emissions from
non-Hermitian valley excitons. The emission is elliptically polarized inside
the exceptional ring (black circle) and linearly polarized outside it. The
two linear polarization directions are parallel at the exceptional ring and gradually
evolve to be orthogonal at large $Q$. (b) Spatial patterns of the exciton
dynamics (green arrows) and their emission properties, when detection energy
is set at the orange (upper), yellow (middle) and green circles (lower) in
the energy spectrum. The angular distributions of emission intensities in
the ballistic transport regime when the polarization direction of detection
fixed at $\hat{x}$ axis are also given. The gray cylinder in the excitonic
spectrum marks the light cone $E/c\sim 10^{-3}K$ and $P=1.5\,\mathrm{meV}$
is used.}
\label{fig3}
\end{figure*}

We now turn to study the valley polarization of these non-Hermitian valley
excitons. The right eigenstates of the Hamiltonian (\ref{H}) read 
\begin{equation}
\left\vert u_{\pm }^{R}\right\rangle \equiv \left( 
\begin{array}{c}
A_{1} \\ 
A_{2}%
\end{array}%
\right) =\left( 
\begin{array}{c}
\pm C_{\pm }\frac{e^{-2i(\varphi -\frac{\pi }{4})}}{\sqrt{4J^{2}Q^{2}-C_{\pm
}^{2}}} \\ 
\frac{2JQ}{\sqrt{4J^{2}Q^{2}-C_{\pm }^{2}}}%
\end{array}%
\right) ,  \label{v}
\end{equation}%
$\allowbreak $where $C_{\pm }=-i\sqrt{4J^{2}Q^{2}-P^{2}K^{2}}\pm PK$ \cite%
{Supp}. The valley polarization is defined as $\eta =\frac{\left\vert
A_{1}\right\vert ^{2}-\left\vert A_{2}\right\vert ^{2}}{\left\vert
A_{1}\right\vert ^{2}+\left\vert A_{2}\right\vert ^{2}}$, which measures the degree of circular polarization in the
photoluminescence experiments \cite{Cui,Hei1,Cao,Sallen,Jones}. In the Hermitian case, the
eigenstates reduce to $\left\vert u_{\pm }^{H}\right\rangle =\frac{1}{\sqrt{2%
}}\left( 
\begin{array}{c}
\pm e^{-2i\varphi } \\ 
1%
\end{array}%
\right) $, which are equal superposition of the two valleys \cite%
{Yu,Yu1,Mac,Lou} and have no valley polarization ($\eta =0$). In
contrast, because of the presence of $\mathcal{PT}$-symmetry breaking, the
non-Hermitian excitonic states $\left\vert u_{\pm }^{R}\right\rangle$ may become valley polarized. As the
steady-state population is dominated by the eigenstate with smaller
decay rate, here we focus on $%
\left\vert u_{+}^{R}\right\rangle $ whose valley polarization is given by \cite{Supp}%
\begin{equation}
\eta =\left\{ 
\begin{array}{c}
\sqrt{1-4J^{2}Q^{2}/P^{2}K^{2}} \ \ \ \ Q<Q_{0} \\ 
\ \ \ \ \ \ \ \ \ \ \ \ \ 0  \ \ \ \ \ \ \ \ \ \ \ \ \ \ \ \ Q>Q_{0}%
\end{array}%
\right. .
\end{equation}%
Fig. \ref{fig2}(c) shows the evolution of $\eta $ as a function of excitonic
momentum $Q$. Remarkably, the state is completely valley polarized at $Q=0$, and becomes partially valley polarized with the
increase of $Q$ in the $\mathcal{PT}$-symmetry broken regime. At the exceptional ring, the valley polarization diminishes completely and keeps zero into
the whole $\mathcal{PT}$-symmetric regime.

For a modest pumping $P>2\,\mathrm{meV}$, the exceptional ring ($Q_{0}>10^{-3}K$)
would locate outside of the light cone ($E/c\sim 10^{-3}K$ within which electron and hole recombine radiatively) and $\eta $
then has an appreciable value. While for a strong pumping of $P=30\,\mathrm{meV}$, $\eta\sim1$ within the whole light cone giving arise to a strong circular polarization in helicity-resolved photoluminescence
measurements (c.f. Fig. \ref{fig2}(c)). In practical experiments, intervalley scattering during exciton formation and difference in exciton decay at the two valleys would result in an effective reduction of exciton pumping \cite{Supp}, which would squeeze the exceptional ring in momentum space and hence reduce the observed valley polarization \cite{Cui,Hei1,Cao,Sallen,Jones}. The consistency between theoretical predictions and experimental observations in turn confirms the validity of our non-Hermitian model.

\textcolor{blue}{Momentum resolved optical emissions}-The non-Hermiticity
also renders the optical emissions from valley excitons an anomalous momentum resolved feature. Arising from the valley-contrasted
optical selection rule \cite{Xu,Yao,Niu3,Yu}, the non-Hermitian valley
excitons couple to photons of the form $\left[ 
\begin{array}{c}
A_{x}e^{i\phi _{x}} \\ 
A_{y}e^{i\phi _{y}}%
\end{array}%
\right] $ in the Jones vector formalism, where $A_{x}e^{i\phi _{x}}=\frac{%
A_{1}+A_{2}}{\sqrt{2\left( \left\vert A_{1}\right\vert ^{2}+\left\vert
A_{2}\right\vert ^{2}\right) }}$ and $A_{y}e^{i\phi _{y}}=i\frac{A_{1}-A_{2}%
}{\sqrt{2\left( \left\vert A_{1}\right\vert ^{2}+\left\vert A_{2}\right\vert
^{2}\right) }}$. The relative phase is \cite{Supp} 
\begin{equation}
\phi \equiv \phi _{y}-\phi _{x}=\left\{ 
\begin{array}{c}
\arctan \frac{C_{+}^{2}-4J^{2}Q^{2}}{-4JQC_{+}\cos 2\varphi } \ \ \ Q<Q_{0} \\ 
\ \ \ \ \ \ \ \ \ \ \ \ 0  \ \ \ \ \ \ \ \ \ \ \ \ \ \ \  \ Q>Q_{0}%
\end{array}%
\right.,
\end{equation}%
which suggests that the optical emission is linearly polarized in the $%
\mathcal{PT}$-symmetry unbroken regime and elliptically polarized in
the broken one \cite{Col}. We note that the Hermitian theory predicts a coupling to
linearly polarized photons of the form $\left[ 
\begin{array}{c}
\cos \alpha _{\pm }^{H} \\ 
\sin \alpha _{\pm }^{H}%
\end{array}%
\right] $ with $\alpha _{\pm }^{H}=\varphi +\left( 1\mp 1\right) \pi /4$ for
the two Hermitian excitonic states $\left\vert u_{\pm }^{H}\right\rangle $. As a
characteristic feature, their linear polarization directions are orthogonal to each other (see
right panel in Fig. \ref{fig1}(b)) \cite{Yu,Yu1,Mac,Lou}. In contrast, our
non-Hermitian theory predicts a much richer emission pattern, as shown in Fig. \ref{fig3}(a).

In the $\mathcal{PT}$-symmetry broken regime, the
non-Hermitian valley excitons couple to elliptically polarized photons of the
form $\left[ 
\begin{array}{c}
A_{x} \\ 
A_{y}e^{i\phi }%
\end{array}%
\right] $ (see \cite{Supp} for details). The ellipticity
of this elliptical polarization is given by $e=2\sqrt{\frac{1}{PK/JQ+2}}$,
which monotonically changes from $0$ at $Q=0$ to $1$ at $Q=Q_0$%
. The limit case of $e=0$ suggests a circularly polarized emission at zero momentum. While the other limit case of $e=1$ suggests a linearly polarized emission at
exceptional ring as we elaborate later. Interestingly,
because the intervalley exchange interaction has a valley-orbital
coupled structure, the major axis of the elliptical polarization is locked
to the excitonic momentum. Explicitly, the angle of
major axis with respect to $\hat{x}$ direction is given by $\alpha =\varphi -\pi /4$ \cite{Supp},
which is totally determined by the excitonic momentum but with a global
phase shift of $-\pi /4$. 

In the $\mathcal{PT}$-symmetric regime, the non-Hermitian excitonic states $\left\vert u_{\pm }^{R}\right\rangle$
become a coherent superposition of the two valleys, which then couple to
linearly polarized photons of the form $\left[ 
\begin{array}{c}
\cos (\alpha _{\pm }^{H}\mp \theta /2) \\ 
\sin (\alpha _{\pm }^{H}\mp \theta /2)%
\end{array}%
\right] $ with $\theta =\arctan \frac{1}{\sqrt{4J^{2}Q^{2}/P^{2}K^{2}-1}}$  \cite{Supp}.
Compared with the Hermitian case, the presence of an additional phase shift 
$\theta $ results in a non-orthogonal feature of the polarization directions
from the two non-Hermitian states (c.f. Fig. \ref{fig3}(a)). Remarkably,
at the exceptional ring, $\theta =\pi /2$ and the polarization directions are both
rotated by $\pi /4$ towards each other and become parallel. This anomalous
polarization distribution is a result of coalescence of eigenstates at
exceptional points, where $\left\vert u_{\pm }^{R}\left(
Q_{0}\right) \right\rangle =\frac{1}{\sqrt{2}}\left( 
\begin{array}{c}
e^{-2i\left( \varphi -\frac{\pi }{4}\right) } \\ 
1%
\end{array}%
\right) $ and both couple to linearly polarized photons of the form $\left[ 
\begin{array}{c}
\cos \left( \varphi -\pi /4\right)  \\ 
\sin \left( \varphi -\pi /4\right) 
\end{array}%
\right] $. At large enough $Q$, $\theta \sim 0$ and the polarization
directions become orthogonal returning to the
Hermitian case.

For a weak optical pumping, orders smaller than the one used in typical experiments for detecting valley polarization, these interesting momentum resolved optical emissions can be detected via using spatial and polarization
resolved photoluminescence measurement (c.f. Fig. \ref{fig3}(b)) \cite{Kulig}. In particular, with the polarization direction of detection fixed
at $\hat{x}$ axis and measured in the $\mathcal{PT}$-symmetric regime with
detection energy $E >E _{\mathrm{EP}}\equiv E
(Q_{0})$ (orange circle in the energy spectrum), the real-space angular distribution of
emission intensity in the ballistic transport regime has a nodal structure (upper panel), with
its major axis having an angle $\alpha ^{\prime }$ relative to the $\hat{x}$%
\ direction that increases monotonically with the decrease of detection
energy $E $. When decreasing $E $ to the exceptional ring with $%
E =E _{\mathrm{EP}}$ (yellow circle), the nodal
structure persists with $\alpha ^{\prime }$ now fixing at $\pi /4$ (middle
panel). Further decreasing detection energy to the $\mathcal{PT}$%
-symmetry broken regime with $E <E _{\mathrm{EP}}$
(green circles), due to the possible coexistence of $\mathcal{PT}$-symmetry
broken and unbroken bands in the light cone, there may exist two emission
patterns (lower panel). One is nodeless with $\alpha' $ still fixing at $\pi
/4$ independent of $E $, attributing to the $\mathcal{PT}$%
-symmetry broken band. The other has a nodal structure with $\alpha ^{\prime
}$ increasing monotonically with the decrease of $E $, attributing
to the $\mathcal{PT}$-symmetry unbroken band. Because excitons at these two
bands have different momentum and hence distinct transport length, they can be separated in real space. We note that the inverse process of above excitonic emissions allows an optical injection of valley excitons with on-demand momentum by using light of selected polarization and energy \cite{Yuhy}.

\begin{figure}[tbp]
\begin{center}
\includegraphics[width=0.475\textwidth]{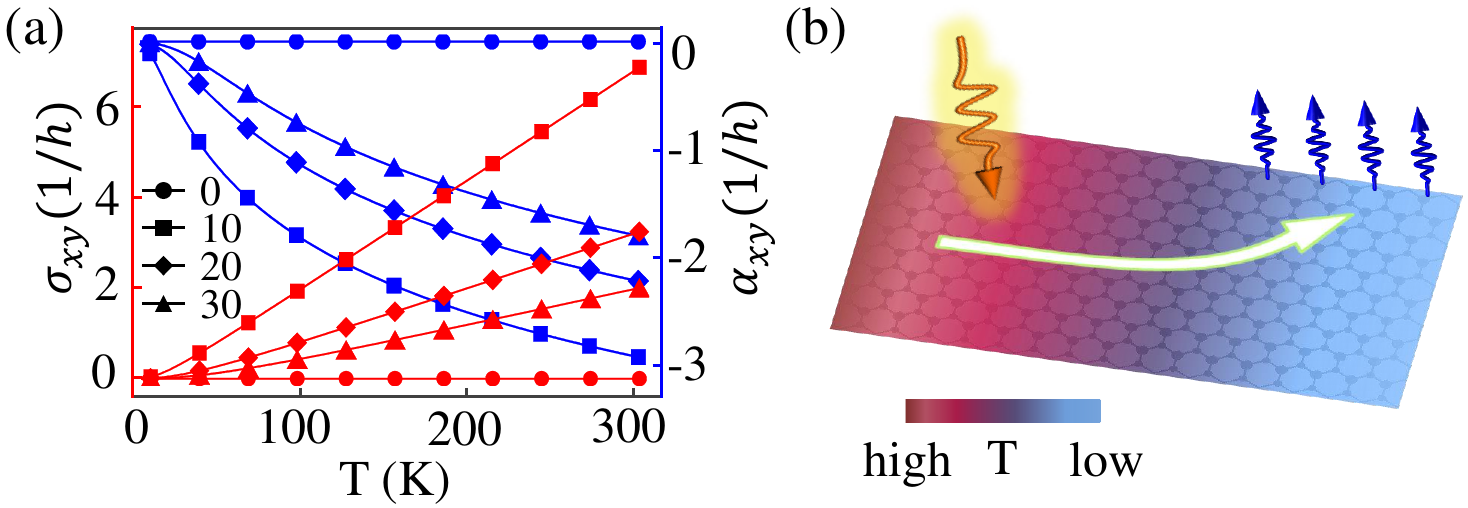}
\end{center}
\caption{(color online) (a) Hall conductivity (red) and Nernst conductivity
(blue) of non-Hermitian valley excitons as a function of temperature for
different pumping $P$ (in unit of \textrm{meV}). $\protect\mu =-1\,$\textrm{%
\ meV} is used. (b) Schematic of an anomalous excitonic Hall
transport, which results in a stronger excitonic emission at the sample
boundary.}
\label{fig4}
\end{figure}

\textcolor{blue}{Anomalous excitonic Hall transport}-The non-Hermiticity
and its resultant $\mathcal{PT}$-symmetry breaking also enables an anomalous excitonic Hall transport that is beyond the
Hermitian predictions. Because a non-Hermitian Hamiltonian has both left and right eigenstates, there are four different definition of Berry curvature \cite{Shen2018}, while only the one defined by right eigenstate contributes to the Hall transport \cite{Supp,Dua,Ila,YX}. Explicitly, the Berry curvature $\Omega _{n}=-\hat{z}\cdot \nabla _{\boldsymbol{Q}}\times 
\mathrm{Im}\left[ \frac{\left\langle u_{n}^{R}\right\vert \partial _{%
\boldsymbol{Q}}\left\vert u_{n}^{R}\right\rangle }{\left\langle
u_{n}^{R}\right. \left\vert u_{n}^{R}\right\rangle }\right] $, which is
required to be zero in the $\mathcal{PT}$-symmetric regime. While in the $\mathcal{PT}$-symmetry broken regime, we find%
\begin{equation}
\Omega _{\pm }=\mp \frac{4J^{2}}{PK\sqrt{P^{2}K^{2}-4J^{2}Q^{2}}},
\end{equation}%
for the two excitonic states $\left\vert u_{\pm
}^{R}\right\rangle $  \cite{Supp}. For pumping $P\sim 10\,\mathrm{meV}$, $
\Omega \sim 10^{4}\,\mathrm{\text{\AA}}^{2}$ in the neighborhood of $Q=0$, which is three orders larger than the
Berry curvature of the electron or hole in monolayer TMDs \cite{Yao}. We
note that $\Omega _{\pm }$ has a singularity at the exceptional ring due to the
coalescence of the non-Hermitian energy spectra. Nevertheless, for steady-state transport,
although $\left\vert u_{+}^{R}\right\rangle $ dominates mostly, $\left\vert
u_{-}^{R}\right\rangle $ takes effect around the exceptional ring through
non-adiabatic contributions, which avoids the singularity in the Hall
current \cite{Supp}. Furthermore, for a strong pumping, the exciton
distribution $f_{n}^{0}\left( \boldsymbol{Q}\right) =\left[ e^{\left( 
\mathrm{Re}[E _{n}]-\mu \right) /k_{B}T}-1\right] ^{-1}$ at the exceptional
ring drops substantially. In the presence of a chemical potential gradient $%
\nabla \mu $ and temperature gradient $\nabla T$, which can be created by a laser beam illuminated on a spot of the sample \cite{Onga2017}, this Berry curvature can lead to an anomalous transverse
excitonic Hall current $j_{H}=\sigma _{H}\nabla \mu -\alpha _{H} k_{B}\nabla T $ \cite{Niu,Niu1,Niu2}, where $\sigma _{H}=-\frac{1}{%
\hbar }\int \frac{d^{2}\boldsymbol{Q}}{\left( 2\pi \right) ^{2}}%
f_{+}^{0}\left( \boldsymbol{Q}\right) \Omega _{+}$ and $\alpha _{H}=\frac{1}{%
\hbar T}\int \frac{d^{2}\boldsymbol{Q}}{\left( 2\pi \right) ^{2}}\Omega _{+}[%
\frac{\mathrm{Re}[E _{+}]-\mu }{k_{B}}\left( f_{+}^{0}\left( \boldsymbol{Q}%
\right) +1\right) +T\log f_{+}^{0}\left( \boldsymbol{Q}\right) ]$ are the
Hall and Nernst conductivity respectively. Fig. \ref{fig4}%
(a) gives the temperature dependence of these two conductivities for different pumping strength. Due to ultrafast exciton formation \cite{Trovatello2020}, a quasi-equilibrium is assumed here \cite{Byrnes2014,Anton2021}, whose derivation from Bose-Einstein distribution would reduce exciton concentration at the $\mathcal{PT}$-symmetry broken regime \cite{Koch2006} and weaken the excitonic Hall effect. This anomalous Hall transport
would lead to a stronger excitonic emission at the sample boundary, as
schematically shown in Fig. \ref{fig4}(b). When the chirality of pumping
light is switched, the Hall current changes sign and a stronger excitonic emission would appear on the opposite boundary. In the presence of disorder, the excitonic Hall effect can still be visualized by a curved transport trajectory of exciton emission via photoluminescence mapping.

\textcolor{blue}{Discussion and conclusions}-For a linearly polarized light driving, the
two valleys are pumped simultaneously with an equal strength and the
Hamiltonian (\ref{H}) reduces to the Hermitian scenario albeit with a trivial
imaginary part. In this case, the system is always in the $\mathcal{PT}$%
-symmetric phase and its eigenstates are a linear superposition of the
two valleys. Accordingly, the optical emission is linearly polarized, which
enables an optical generation of excitonic valley coherence as observed in experiments 
\cite{Jones,Wang2015}. We note that a full non-Hermitian modeling of valley excitons requires \textit{ab initio} simulations including microscopic informations \cite{Rohlfing1998,Rohlfing2000,Deslippe2012,Cohen2016}. The non-Hermitian excitonic transport can be modeled using non-equilibrium quantum theory with properly incorporated pumping and decay \cite{Evers2013,Evers2020,Evers2024}. Moreover, non-Hermiticity can also appear due to nonreciprocal hopping between the two valleys arising from valley decoherence \cite{Supp}.

In summary, our non-Hermitian valley-exciton theory revealed a $\mathcal{PT}$-symmetry breaking induced
novel valley-polarized excitonic state with elliptically polarized optical emission, which is the intrinsic mechanism that obstructs the observation of intervalley excitonic coherence effect in helicity-resolved optical
experiments. At large excitonic momenta, $\mathcal{PT}$-symmetry
restores and the optical emissions become linearly polarized with their polarization directions having a non-orthogonal feature. The non-Hermiticity also allows a non-zero Berry curvature for valley excitons, which enables an anomalous excitonic Hall transport.  Our work not only provides an alternative dynamical understanding on intriguing valley excitonic phenomena in 2D TMDs, but also suggests a highly tunable low-dimensional solid-state platform to explore non-Hermitian physics without Hermitian counterparts.

\begin{acknowledgments}
We thank Wang Yao for helpful discussion. This work was supported by the National Key Research and Development Program of Ministry of Science and Technology (2022YFA1204700, 2021YFA1200503), the National Natural Science Foundation of China (12374178), the Science Fund for Distinguished Young Scholars of Hunan Province (2022J10002), and the Fundamental Research Funds for the Central Universities from China.
\end{acknowledgments}

\end{document}